\PassOptionsToPackage{unicode}{hyperref}
\PassOptionsToPackage{hyphens}{url}
\PassOptionsToPackage{dvipsnames,svgnames,x11names}{xcolor}
\documentclass[
  british,
  10pt,
  a4paper,
]{article}
\usepackage{xcolor}
\usepackage[margin=24mm]{geometry}
\usepackage{amsmath,amssymb}
\usepackage{iftex}
\ifPDFTeX
  \usepackage[T1]{fontenc}
  \usepackage[utf8]{inputenc}
  \usepackage{textcomp} 
\else 
  \usepackage{unicode-math} 
  \defaultfontfeatures{Scale=MatchLowercase}
  \defaultfontfeatures[\rmfamily]{Ligatures=TeX,Scale=1}
\fi
\usepackage{lmodern}
\ifPDFTeX\else
\fi
\IfFileExists{upquote.sty}{\usepackage{upquote}}{}
\IfFileExists{microtype.sty}{
  \usepackage[]{microtype}
  \UseMicrotypeSet[protrusion]{basicmath} 
}{}
\makeatletter
\@ifundefined{KOMAClassName}{
  \IfFileExists{parskip.sty}{%
    \usepackage{parskip}
  }{
    \setlength{\parindent}{0pt}
    \setlength{\parskip}{6pt plus 2pt minus 1pt}}
}{
  \KOMAoptions{parskip=half}}
\makeatother
\usepackage{longtable,booktabs,array}
\usepackage{caption}
\usepackage{calc} 
\usepackage{etoolbox}
\makeatletter
\patchcmd\longtable{\par}{\if@noskipsec\mbox{}\fi\par}{}{}
\makeatother
\IfFileExists{footnotehyper.sty}{\usepackage{footnotehyper}}{\usepackage{footnote}}
\makesavenoteenv{longtable}
\ifLuaTeX
\usepackage[bidi=basic,shorthands=off]{babel}
\else
\usepackage[bidi=default,shorthands=off]{babel}
\fi
\ifLuaTeX
  \usepackage{selnolig} 
\fi
\providecommand{\tightlist}{%
  \setlength{\itemsep}{0pt}\setlength{\parskip}{0pt}}
\usepackage{microtype}
\usepackage{xurl}
\usepackage{booktabs}
\usepackage{mathtools}
\usepackage{bookmark}
\IfFileExists{xurl.sty}{\usepackage{xurl}}{} 
\makeatletter
\@ifundefined{xmpquote}{}{}
\makeatother
\hypersetup{
  pdftitle={Certified exact identification of the I3322 quantum value},
  pdfauthor={Artus Krohn-Grimberghe (Percivio Ltd.)},
  pdflang={en-GB},
  colorlinks=true,
  linkcolor={blue},
  filecolor={Maroon},
  citecolor={Blue},
  urlcolor={blue},
  pdfcreator={LaTeX via pandoc}}

\title{Certified exact identification of the I3322 quantum value}
\author{Artus Krohn-Grimberghe (Percivio Ltd.)}
\date{30 August 2026}

\begin{document}
\maketitle

\textbf{Preprint.} The exact finite certificate chain has been replayed
on its declared surfaces. The functional-analytic optimizer-to-fibre
passage and the dimension-free operator response lemma are written
proofs rather than full formalizations. A read-only automated
adversarial review by a model from a second vendor found no mathematical
error, but it is not human peer review. Its three major presentation and
dependency findings are repaired in this version; the complete verdict
and its provenance are included in the verification deposit described in
Appendix B.

\subsection{Abstract}\label{abstract}

We determine the quantum value of the three-setting, two-outcome Bell
functional I3322. The value is the unique intersection \(q_\dagger\) of
an analytic A-high stable curve of the exact Jacobi characteristic map
with a central reverser curve. A computer-assisted proof isolates this
intersection in the interval

\begin{center}
\resizebox{0.98\linewidth}{!}{$
\begin{aligned}
0.2508753845139765356173366109459512989013219739714764755
&<q_\dagger\\
&<0.2508753845139765356173366109459512989013219739714764756.
\end{aligned}
$}
\end{center}

We prove that the supremum over finite-dimensional tensor-product
strategies, the unrestricted spatial value, and the commuting-operator
value all equal \(q_\dagger\). No finite-dimensional strategy attains
the value, whereas a normal infinite-dimensional spatial strategy and a
commuting GNS strategy do. The proof combines an exact Bellman/path
variational reduction, extraction of a positive bilateral calibrated
Jacobi fibre from a maximizing weak-star state, a reflected-fibre
central-incidence theorem, and interval certificates for the unique
stable/central intersection. A separate operator-positive
three-remainder decomposition gives the commuting upper bound. We
explain why an earlier exact amplitude-gap countercertificate remains
valid but does not enter either inequality in the present proof. The
constant \(q_\dagger\) is characterized analytically; no algebraic
equation or claim about its arithmetic nature is made.

\subsection{1. Introduction}\label{introduction}

Collins and Gisin introduced I3322 as the unique new facet, up to the
usual equivalences, in the bipartite three-setting, two-outcome Bell
scenario {[}1{]}. It is the simplest such inequality beyond CHSH in its
number of measurement settings. In contrast to CHSH, increasingly large
finite-dimensional strategies improve its known quantum violation. Pál
and Vértesi constructed an explicit sequence of strategies with growing
local dimension and conjectured that finite dimension cannot attain the
true value {[}2{]}. Gigena and Kaniewski later placed I3322 in the
numerically treated branch of a wider three-parameter family of Bell
functionals {[}3{]}.

Two proof-carrying 2026 developments sharpened the numerical and
structural picture. Mghirbi certified a rational enclosure of width
below \(10^{-9}\), without claiming the exact value or attainment
{[}4{]}. Douglas proved equality of the tensor and commuting variational
values, finite-dimensional nonattainment, normal spatial attainment, and
logarithmic dimension complexity, while explicitly leaving the exact
value within a certified window open {[}5{]}. A contemporaneous preprint
by Connor independently claims the qualitative infinite-dimensional
attainment and nonattainment result through a compact variational
problem over probability measures {[}6{]}. We therefore claim no
priority for those qualitative statements.

The contribution of this paper is the exact identification of the common
value with one analytically characterized constant: the unique certified
intersection \(q_\dagger\) of the A-high stable curve and the central
reverser curve, together with the width-\(10^{-55}\) rational enclosure
in the abstract. The distinction matters because a previous attempted
identification of the global variational value with a shooting constant
was decertified by an exact amplitude mismatch even though its local
position and ratio coordinates matched.

This paper closes the remaining value-identification problem. The
decisive change is not a more accurate shooting computation. It is a
change in proof architecture:

\begin{enumerate}
\def\labelenumi{\arabic{enumi}.}
\tightlist
\item
  construct the lower bound from a bilateral Jacobi eigenvector, which
  needs only the physical ratio recurrence;
\item
  construct the upper implication from the genuinely normalized
  maximizing Bellman/KKT state;
\item
  extract one atomic conditional response orbit from that state;
\item
  compare the extracted orbit with its separately constructed
  reflection, avoiding any assumption that a conditional component is
  itself party-self-dual;
\item
  use exact depth and parity certificates to force the orbit through the
  unique central intersection.
\end{enumerate}

This separates the two normalization problems that were conflated in the
decertified construction.

\subsubsection{1.1 Main theorem}\label{main-theorem}

Let \(\Omega_\mathrm{fd}\) be the supremum over finite-dimensional
tensor-product strategies, \(\Omega_\mathrm{sp}\) the supremum over
arbitrary spatial tensor-product strategies, and \(\Omega_\mathrm{qc}\)
the supremum over commuting-operator strategies for the normalization in
Section 2.

\textbf{Theorem A.} There is a uniquely characterized real number
\(q_\dagger\in(1/4,1/3)\) such that

\[
\boxed{
\Omega_\mathrm{fd}=\Omega_\mathrm{sp}=\Omega_\mathrm{qc}=q_\dagger.
}
\]

The finite-dimensional supremum is not attained in any finite local
dimensions. The spatial value is attained by a normal vector state on an
infinite-dimensional tensor product, and the commuting value is attained
by a state of the universal commuting \(C^*\)-algebra and its GNS
representation.

\textbf{Theorem B.} The constant \(q_\dagger\) is the unique zero in the
certified box of the two-component analytic residual \(R_{15}\) defined
in Section 4, and satisfies the width-\(10^{-55}\) enclosure displayed
in the abstract.

Theorem B is an exact analytic characterization. The finite recurrence
uses only rational operations and positive square roots, but the stable
curve is selected by an infinite-iteration condition. We do not know
whether \(q_\dagger\) is algebraic.

\subsubsection{1.2 What is new and what is
inherited}\label{what-is-new-and-what-is-inherited}

The Pál--Vértesi block family and the infinite-dimensional advantage are
not new. Nor do we claim priority for the common Bellman/path value,
finite-dimensional nonattainment, the existence of a normal spatial
maximizer, or the logarithmic dimension law
\(D(\varepsilon)=\Theta(\log(1/\varepsilon))\) {[}5,6{]}. The new
contribution is the exact identification of that common value with the
unique certified stable/central intersection, together with a proof
architecture that survives the historical amplitude countercertificate.
We also record a compact commuting upper-bound argument specialized to
the same exact constant.

\subsection{2. Bell functional and
models}\label{bell-functional-and-models}

Let \(A_1,A_2,A_3\) and \(B_1,B_2,B_3\) be projections. In the commuting
model every Alice projection commutes with every Bob projection. We use
the Bell element

\[
\begin{aligned}
\mathcal B={}&-A_2-B_1-2B_2
 +A_1B_1+A_1B_2+A_2B_1+A_2B_2\\
&-A_1B_3+A_2B_3-A_3B_1+A_3B_2.
\end{aligned}
\tag{2.1}
\]

Local outcome flips convert (2.1) to the equivalent reflection
convention used in the scalar recurrence. They do not change any of the
three quantum values.

Define

\[
b(x)=\frac{\sqrt{1-x^2}}2,
\qquad
d(y,x)=yx+\frac{y-x}{2}-1.
\tag{2.2}
\]

The alternating two-projection decomposition turns every finite scalar
label path \(c_0,\ldots,c_n\in[-1,1]\) into a finite Jacobi carrier with
diagonal entries \(d(c_{j-1},c_j)\) and off-diagonal entries \(b(c_j)\).
Conversely, the Jordan blocks of a finite I3322 strategy split into such
carriers.

Let \(S_\mathrm{path}\) be the supremum of the top eigenvalues of all
finite carriers. A Schur-pivot recursion gives the Bellman variational
value \(S_\mathrm{Bellman}\): it is the least \(q\) for which positive
terminal storages exist with

\[
d(x,u)+\frac{b(x)^2}{g(x)}+g(u)\le q
\quad(-1\le x,u\le1).
\tag{2.3}
\]

The finite reduction (premise {[}P1{]}) proves

\[
\Omega_\mathrm{fd}=S_\mathrm{path}=S_\mathrm{Bellman}.
\tag{2.4}
\]

Exact finite strategies and the crossed Bellman certificate give

\[
\frac14<L_{1001}<L_9\le S_\mathrm{Bellman}\le U_{15}<\frac13,
\tag{2.5}
\]

with the exact rational values of \(L_{1001}\), \(L_9\) and \(U_{15}\)
recorded under premise {[}P2{]}. The two inner inequalities are
inherited as non-strict; both become strict once Theorem A is proved,
since \(L_9<q_\dagger<U_{15}\) with margins exceeding
\(2.6\times10^{-40}\) and \(8.9\times10^{-13}\) (Theorem 4.1).

Equations (2.4) and (2.5) are the first two inherited results. Every
inherited lemma used in this paper is stated in Appendix C in the exact
form in which it is invoked, labelled {[}P1{]}--{[}P11{]}, and bound to
its source document by SHA-256. The labels are cited at the point of
use.

\subsection{3. The characteristic
recurrence}\label{the-characteristic-recurrence}

For a state \(z=(x,y,r)\), define

\[
r^+=\frac{q-d(y,x)-b(x)/r}{b(y)},
\tag{3.1}
\]

\[
y^+=\frac12-\frac{1+2x}{2(r^+)^2}
     +\frac{y}{\sqrt{1-y^2}\,r^+},
\tag{3.2}
\]

and

\[
T_q(x,y,r)=(y,y^+,r^+).
\tag{3.3}
\]

If \(z_j=(a_{j-1},a_j,r_{j-1})\) and
\(r_{j-1}=\lambda_j/\lambda_{j-1}\), then (3.1) is exactly the Jacobi
eigenvalue equation divided by \(\lambda_j\). Equation (3.2) is the
label stationarity equation.

The positive A-high fixed branch is

\[
z_*(q)=(C(q),C(q),R(q)),
\tag{3.4}
\]

where

\[
C(q)=\sqrt{\frac{5+4q+\sqrt{16q^2+24q-7}}8},
\tag{3.5}
\]

\[
R(q)=\frac{2C(q)+1}{2C(q)-1}
      \sqrt{\frac{1-C(q)}{1+C(q)}}.
\tag{3.6}
\]

On the interval (2.5), \(0<R(q)<1\). The fixed point has a unique
phase-normalized one-dimensional analytic stable curve \(W_q(t)\). The
computer-assisted stable-manifold proof treats separately the constant
mode, the augmented eigenvector/multiplier/phase mode, and all modes of
order at least two. It certifies, uniformly in \(q\in[L_9,U_{15}]\), a
complex-disk error below \(5\times10^{-48}\) for the degree-eight
polynomial chart used by the later finite covers (premise {[}P8{]}).

The edge reverser is state-dependent:

\[
\mathscr R_q(x,y,r)=(-y,-x,1/r^+),
\tag{3.7}
\]

and satisfies

\[
\mathscr R_q^2=\mathrm{id},
\qquad
\mathscr R_qT_q\mathscr R_q=T_q^{-1}.
\tag{3.8}
\]

Using \(1/r\) in place of \(1/r^+\) in (3.7) is generally wrong. The
reverser identities are premise {[}P7{]}.

\subsection{\texorpdfstring{4. Definition and isolation of
\(q_\dagger\)}{4. Definition and isolation of q\_\textbackslash dagger}}\label{definition-and-isolation-of-q_dagger}

Starting from \(W_q(t)\), apply the exact inverse recurrence fifteen
times:

\[
(p_{15},x_{15},r_{15})=T_q^{-15}W_q(t).
\tag{4.1}
\]

The central curve is

\[
r_c(q,p)=\frac{q+p^2-p+1}{\sqrt{1-p^2}},
\tag{4.2}
\]

\[
x_c(p,r)=\frac12-\frac{1-2p}{2r^2}
              +\frac{p}{\sqrt{1-p^2}\,r}.
\tag{4.3}
\]

Define the two-component residual

\[
R_{15}(q,t)=
\begin{pmatrix}
r_{15}-r_c(q,p_{15})\\
x_{15}-x_c(p_{15},r_{15})
\end{pmatrix}.
\tag{4.4}
\]

The certified Poincaré--Miranda box supplies a zero. Uniqueness rests on
the following capture theorem. It is stated in full because it is the
step that turns a local isolation into a global characterization, and
because its domain is exactly what the rest of the proof must hit.

\textbf{Theorem 4.1 (global capture: annulus \(\to\) strip \(\to\)
box).} Let \(q\in[L_9,U_{15}]\). Let

\[
A_+=[\,1.71\times10^{-12},\,10^{-11}\,],\qquad
A_-=[\,-10^{-11},\,-1.71\times10^{-12}\,]
\tag{4.5}
\]

be the two closed fundamental annuli of the phase-fixed chart. On the
whole \(q\)-interval the stable multiplier satisfies
\(0.1712920057349488984\ldots<\lambda_s(q)<0.1712920057351056990\ldots\),
so every nonzero orbit of the local stable manifold has a representative
in \(A_+\cup A_-\) (premise {[}P8{]}). Suppose \(t\in A_+\cup A_-\),
that \(R_n(q,t)=0\) for some inverse depth \(n\ge0\), and that the
inverse orbit through \(W_q(t)\) is a complete positive A-high fibre.
Then:

\begin{enumerate}
\def\labelenumi{(\roman{enumi})}
\tightlist
\item
  (annulus \(\to\) strip) \(n=15\), \(t\in A_+\), and the point lies in
  the retained strip
\end{enumerate}

\begin{equation}
\resizebox{0.98\linewidth}{!}{$
[\,q_-,U_3\,]\times[\,s_-,s_+\,],\qquad
q_-=\frac{6271884612849413390433415273648782472283049349286911889555958159881884999837}{25\cdot10^{76}},
\qquad
U_3=\frac{250875384513991691}{10^{18}}
$}
\tag{4.6}
\end{equation}

\[
s_-=\frac{420792173}{8\cdot10^{19}},\qquad
s_+=\frac{2103961197}{4\cdot10^{20}},
\]

where \(s\) is the coordinate of the same point in the retained stable
parameterization, with \(|s-t|<1.318\times10^{-21}\).

\begin{enumerate}
\def\labelenumi{(\roman{enumi})}
\setcounter{enumi}{1}
\tightlist
\item
  (strip \(\to\) box) Every zero of the retained residual in the strip
  (4.6) lies in the isolator
\end{enumerate}

\[
|q-Q_\mathrm{ref}|\le10^{-38},\qquad|\eta|\le10^{-48},
\tag{4.7}
\]

with
\(Q_\mathrm{ref}=0.25087538451397653561733661094595129890132197397147647558\ldots\)
and \(\eta\) the bridged local coordinate, and the isolator contains a
zero by Poincaré--Miranda.

\begin{enumerate}
\def\labelenumi{(\roman{enumi})}
\setcounter{enumi}{2}
\tightlist
\item
  (uniqueness) On (4.7) the interval Jacobian of the residual map has
  determinant in
  \([\,7.9596388720704975,\,7.9596388722636150\,]\times10^{14}\), so the
  map is injective there and the zero is unique. The unique zero
  \((q_\dagger,t_\dagger)\) satisfies
  \(q_\dagger-L_9>2.65\times10^{-40}\) and
  \(U_{15}-q_\dagger>8.98\times10^{-13}\).
\end{enumerate}

\emph{Proof of (i).} The statement is a finite directed interval cover
of the product box \([L_9,U_{15}]\times(A_+\cup A_-)\) over all inverse
depths. The tail classification {[}P6{]} confines the inverse orbit to
the positive A-high entrance, and completeness of the fibre supplies
positivity of every label and ratio along it. On \(A_+\): depths
\(0\le n\le14\) are excluded on the whole product box by a directed sign
of one residual component; depth 16 is excluded on each of 64
\(t\)-cells; at depth 15 an initial partition into 64 cells excludes 63,
and the survivor is reduced by seven further subdivisions to the
buffered interval

\[
5.2599021645\times10^{-12}\le t\le5.2599029905\times10^{-12},
\tag{4.8}
\]

on which the high-\(q\) strip \(U_3\le q\le U_{15}\) is excluded by the
second residual component; at depth 17 the label \(x_{17}\) is negative
on all 64 cells, so every depth \(n\ge17\) violates positivity. On
\(A_-\): depths \(0\le n\le13\) are excluded, and at depths 14 and 15
every box reaches \(|x|=1\), 74 boxes directly and 54 through the
cancellation-preserving dichotomy \(x_{15}>1\). The fixed point \(t=0\)
is excluded because its residual is separated from zero. Altogether 358
terminal boxes are certified through 5106 directed denominator checks
with minimum denominator margin \(0.2104573707778315\ldots\). Combining
(4.8) with \(|s-t|<1.318\times10^{-21}\) and \(q_-<L_9\) places the
surviving point in (4.6). The cover, its ledger and its checker are
premise {[}P9{]}.

\emph{Proof of (ii) and (iii).} Strip-to-box confinement, the coordinate
bridge and the Poincaré--Miranda isolator are the members
\texttt{global\_exclusion\_checker.py},
\texttt{coordinate\_bridge\_checker.py} and
\texttt{local\_isolator\_checker.py} of the exact-checker archive of
premise {[}P9{]}, replayed assertion-free in ordinary and optimized
modes by \texttt{verify\_hardened\_r15.py}; the determinant hull is
reproduced by the independent Jacobian reconstruction
\texttt{local\_injectivity\_audit.py}. Appendix C records the digests.
\(\square\)

Theorem 4.1 defines the unique zero \((q_\dagger,t_\dagger)\). Sections
5--7 show that the maximizing fibre lands in its hypotheses: the fibre
is complete, positive and A-high (premise {[}P6{]}), its chart
coordinate has a representative in \(A_+\cup A_-\) (premise {[}P8{]}),
and its central incidence is a zero of \(R_{15}\) (Section 7).

An independent 768-bit dyadic implementation encloses the
\(q\)-coordinate in the width-\(10^{-55}\) interval of the abstract. The
implementation imports no floating-point root as an axiom; it uses the
hash-bound rational stable-graph coefficient table together with the
previously certified analytic remainder.

\subsection{5. The lower inequality}\label{the-lower-inequality}

At the isolated zero, all fifteen inverse boxes are physical: both label
coordinates lie in \((0,1)\), every denominator is separated from zero,
and every amplitude ratio is positive. The forward tail stays on the
A-high stable curve and its ratios converge to \(R(q_\dagger)<1\).

One forward step from a reverser-fixed bond state lies on the central
curve:

\[
T_q\!\left(-p,p,\frac1{r_c(q,p)}\right)
=\left(p,x_c(p,r_c(q,p)),r_c(q,p)\right).
\tag{5.1}
\]

Equation (5.1), the bond-midpoint identity of premise {[}P7{]}, supplies
the reciprocal seam. Choose one positive central amplitude and propagate
it with the ratios. Reflection gives the opposite tail. The resulting
vector satisfies

\[
\lambda_j>0,
\qquad \lambda\in\ell^2(\mathbb Z),
\qquad J\lambda=q_\dagger\lambda.
\tag{5.2}
\]

For the compression \(J_N=P_NJP_N\),

\[
\frac{\langle P_N\lambda,J_NP_N\lambda\rangle}
     {\|P_N\lambda\|^2}
\longrightarrow q_\dagger.
\tag{5.3}
\]

Every finite carrier embeds in a finite-dimensional tensor-product I3322
strategy. Hence

\[
q_\dagger\le\Omega_\mathrm{fd}.
\tag{5.4}
\]

This argument uses no globally glued Bellman storage.

\subsection{6. From a maximizing state to a complete calibrated
fibre}\label{from-a-maximizing-state-to-a-complete-calibrated-fibre}

We now work at \(q=\Omega_\mathrm{fd}\), without assuming its value.

For each fixed finite dimension compactness supplies an exact maximizer.
Helstrom response permits kernels, and direct-summing a strategy with
its party reflection gives a self-dual maximizing sequence. Weak-star
compactness of states on the universal minimal tensor algebra supplies a
maximizing state and its GNS representation.

The terminal-pivot storages are uniformly bounded, concave and
Lipschitz. Arzelà--Ascoli gives a limiting storage \(g\). The bound
\(\Omega_\mathrm{fd}<1/3\) gives a strict endpoint barrier, so \(g>0\)
on the closed interval. The positive Bellman weld remainders therefore
converge in operator norm and annihilate the maximizing vector (premise
{[}P3{]}).

The joint scalar spectral measure is carried by the one-to-one strictly
increasing zero graph \(Z=\{(P(u),u)\}\), with no mass at the endpoints
(premise {[}P4{]}). Zero-slack localization yields two decreasing
response involutions and exact positive Radon--Nikodym cocycles (premise
{[}P5{]}). Their increasing product generates a countable
\(\mathbb Z\)-action. A rational-cut construction supplies a Borel
transversal; adjoining one response involution gives a transversal for
the full infinite-dihedral relation. The fixed part has zero maximizing
mass, because a fixed orbit would imply

\[
\frac14-\Omega_\mathrm{fd}
=\left(\sin m-\frac12\right)^2+\sin^2h\ge0,
\tag{6.1}
\]

contrary to (2.5).

Disintegrate the maximizing scalar measure over the orbit quotient. For
almost every conditional probability, uniqueness of disintegration
transfers both cocycles. A probability on a countable orbit has an atom;
positive finite cocycles propagate positive mass to every orbit point.
Interleaving the two parities and taking square roots of the atomic
masses gives

\[
\lambda_j>0,
\qquad
\sum_j\lambda_j^2=1,
\qquad
\frac{\lambda_{j+1}}{\lambda_j}=\frac{g(c_j)}{b(c_j)}.
\tag{6.2}
\]

Bellman equality gives the literal Jacobi calibration

\[
J\lambda=\Omega_\mathrm{fd}\lambda.
\tag{6.3}
\]

This conclusion concerns a selected conditional orbit. It does not
assert that the original GNS measure has a global atom, that the
conditional is a positive-mass reducing summand, or that every
conditional orbit is good.

Strict ordering and bilateral square summability exclude constant
orbits. The exact endpoint, singular, recurrent, B, mixed and A-low
analyses leave only the A-high heteroclinic tails (premise {[}P6{]}).
Thus the selected fibre enters the certified A-high fundamental annulus
(premise {[}P8{]}).

\subsection{7. Reflected-fibre central
incidence}\label{reflected-fibre-central-incidence}

The selected conditional fibre need not itself be party-self-dual. This
was the central quantifier obstruction in earlier proof attempts.

Construct the reflected fibre separately. With

\[
z_j=(a_{j-1},a_j,r_{j-1}),
\qquad T_qz_j=z_{j+1},
\]

its states satisfy

\[
z_k^\sharp=\mathscr R_qz_{1-k}.
\tag{7.1}
\]

Choose positive-annulus representatives \(W_q(t)=z_j\) and
\(W_q(s)=z_k^\sharp\). Applying the reverser to the latter produces a
point of the original orbit. Therefore, for some positive integer \(N\),

\[
C_{N-n}(q,t)=\mathscr R_qC_n(q,s),
\tag{7.2}
\]

where \(C_n=T_q^{-n}W_q\). No componentwise invariance was assumed.

The exact depth signs

\[
x_{15}>0,
\qquad x_{17}<0,
\qquad O_{17}=x_{17}+y_{17}<0
\tag{7.3}
\]

and strict inverse-depth ordering imply \(N\in\{32,33,34\}\).

\begin{itemize}
\tightlist
\item
  At \(N=32\), the PS32 divided-difference certificate forces the two
  phases to coincide. The midpoint is reverser-fixed, hence
  bond-centred.
\item
  At \(N=33\), the mixed-depth cover excludes the nonzero sector and the
  SC16 certificate excludes the both-zero sector.
\item
  At \(N=34\), (7.3) contradicts the required reflected sum.
\end{itemize}

Every extracted complete fibre is therefore bond-centred. Applying one
forward step to its reverser-fixed midpoint gives

\[
R_{15}(\Omega_\mathrm{fd},t)=0.
\tag{7.4}
\]

The unique-zero theorem of Section 4 forces

\[
\Omega_\mathrm{fd}=q_\dagger.
\tag{7.5}
\]

Together with (5.4), this proves the finite-dimensional value identity.

\subsection{8. Finite-dimensional
nonattainment}\label{finite-dimensional-nonattainment}

Suppose a finite-dimensional strategy attained \(q_\dagger\). It enters
the construction of Section 6 legitimately: a mixed attaining state may
be replaced by a vector in the top eigenspace of its Bell operator,
binary effects admit a finite simultaneous Naimark dilation, and after
equal padding the coherent direct sum of the strategy with its party
flip is a party-self-dual exact maximizer whose padded copies form a
constant maximizing sequence. Its weak-star limit is the strategy
itself, so the attained storage \(g\), the zero graph \(Z=\{(P(u),u)\}\)
and the two response transport laws of Section 6 (premises
{[}P3{]}--{[}P5{]}) hold for it verbatim.

Let \(m=\mu_U\) be the target spectral marginal. Its GNS representation
is finite-dimensional, so \(m\) has finite atomic support \(\Sigma\).
The endpoints are unoccupied: at \(t=\pm1\) one has
\(p(t)=b(t)^2/g(t)=0\) while \(g(\pm1)>0\), so an occupied endpoint
would violate the contact relation \(p(t)=g(-t)\). Hence
\(\Sigma\subset(-1,1)\).

\textbf{Totality of the response maps on \(\Sigma\).} The response maps
of Section 6 are

\[
\alpha(u)=P^{-1}(-P(u)),\qquad\sigma(u)=-u.
\tag{8.1}
\]

That both are defined at every occupied atom requires proof. The
identities \(P(Y)=-Y\) and \(P(-P(u))=-u\) alone do not give it: the
two-point set \(Y=\{1/3,2/3\}\) with \(P(1/3)=-2/3\), \(P(2/3)=-1/3\) is
strictly increasing and satisfies both identities, yet \(-P(u)\) is
positive for both points and lies outside the domain
\(P(Y)=\{-2/3,-1/3\}\) of \(P^{-1}\), so \(\alpha\) is defined nowhere
on \(Y\). This example was recorded when the historical global-closure
argument was decertified (Appendix C.2); it is why Section 6 works on a
conull set after removing exceptional domains, and a proof that treats
\(\alpha\) as a bijection of the finite support without further argument
is incomplete.

On a finite atomic support the obstruction disappears, because the
transport laws are measure identities with strictly positive finite
densities. The reflection laws of Section 6 read

\[
(-\mathrm{id})_*\mu_X=r_A^2\,\mu_X,\qquad
(-\mathrm{id})_*\mu_U=r_B^2\,\mu_U,
\tag{8.2}
\]

\[
r_A(t)=\sqrt{\frac{g(-t)}{g(t)}},\qquad
r_B(t)=\sqrt{\frac{g(t)}{g(-t)}},\qquad
r_A(t)\,r_A(-t)=r_B(t)\,r_B(-t)=1,
\tag{8.3}
\]

and both densities are finite and strictly positive on \((-1,1)\)
because \(g>0\) on \([-1,1]\). Evaluating (8.2) on singletons gives, for
every occupied \(u\in\Sigma\) and every occupied \(x=P(u)\),

\[
\mu_U(\{-u\})=r_B(u)^2\,\mu_U(\{u\})>0,\qquad
\mu_X(\{-x\})=r_A(x)^2\,\mu_X(\{x\})>0.
\tag{8.4}
\]

So the supports of both marginals are symmetric under negation. Since
\(\mu_X=P_*\mu_U\) and \(P\) is injective on \(\Sigma\), the occupied
\(X\)-atoms are exactly \(P(\Sigma)\); hence \(-u\in\Sigma\) and
\(-P(u)\in P(\Sigma)\) for every \(u\in\Sigma\). This is precisely the
statement that \(\sigma\) and \(\alpha\) are defined on all of
\(\Sigma\) and map \(\Sigma\) into itself. (In the two-point example,
(8.4) fails: \(-1/3\notin Y\).) Both maps are involutions,
\(\alpha^2(u)=P^{-1}(P(u))=u\), hence bijections of \(\Sigma\), and both
are strictly decreasing because \(P\) is strictly increasing. In orbit
form the laws read \(\alpha_*m=r_A(P(\cdot))^2m\) and
\(\sigma_*m=r_B^2m\); since the densities are reciprocal under the
corresponding involution, singleton evaluation gives the oriented
singleton laws

\[
m(\{s(u)\})=q_s(u)\,m(\{u\})>0,\qquad s\in\{\alpha,\sigma\},\qquad
q_\alpha(u)=r_A(P(u))^2,\quad q_\sigma(u)=r_B(u)^2.
\tag{8.5}
\]

\textbf{Finite order forces the identity.} The composition
\(\tau=\alpha\sigma\) is a strictly increasing bijection of the finite
totally ordered set \(\Sigma=\{u_1<\dots<u_N\}\). Surjectivity forces
\(\tau(u_1)=u_1\), and induction fixes every later point, so
\(\tau=\mathrm{id}\). (The shipped finite checkers exercise this lemma
only for supports of size at most eight; they are structural controls,
not its proof.) Hence \(\alpha(u)=\sigma(u)=-u\) on \(\Sigma\), that is,
\(P(-u)=-P(u)\) for every occupied \(u\). Put \(x=P(u)\). The two laws
(8.5) have the same target atom \(-u\), so positivity gives
\(\rho:=r_A(x)=r_B(u)>0\), and the derived central pair \((-x,-u)\) is a
zero pair. It must be distinguished from the party reflection
\((-u,-x)\): at \(x=1/2\), \(u=1/4\) the values \(d(-u,-x)=-3/4\) and
\(d(-x,-u)=-1\) differ, and substituting the wrong pair invalidates the
elimination below.

\textbf{Elimination.} Zero localization, \(p(x)=g(-x)\) and
\(p(-u)=g(u)\), together with \(\rho=r_A(x)=r_B(u)\), gives
\(g(-x)=\rho\,b(x)\), \(g(x)=b(x)/\rho\), \(g(u)=\rho\,b(u)\),
\(g(-u)=b(u)/\rho\). Bellman contact at \((x,u)\) and at \((-x,-u)\)
then reads, with \(\Lambda=b(x)+b(u)\), \(S=q_\dagger-xu+1\) and
\(\delta=(x-u)/2\),

\[
S-\delta=\rho\Lambda,\qquad S+\delta=\Lambda/\rho,
\tag{8.6}
\]

so \(S>0\) and \(S^2=\Lambda^2+\delta^2\). With \(x=\cos\alpha\),
\(u=\cos\beta\), \(m=(\alpha+\beta)/2\), \(h=(\alpha-\beta)/2\) one has
\(\Lambda=\sin m\cos h\), \(\delta=-\sin m\sin h\), hence \(S=\sin m\),
and \(xu=1-\sin^2m-\sin^2h\); therefore

\[
q_\dagger=xu-1+\sin m
=\frac14-\left(\sin m-\frac12\right)^2-\sin^2h\le\frac14,
\]

which is (6.1). This contradicts the certified strict lower bound
\(q_\dagger>1/4\) (premise {[}P2{]}). Hence no finite local dimensions
attain the supremum. The ceiling is sharp at \(q=1/4\)
(\(x=u=\pm\sqrt3/2\), \(\rho=1\)), so the strict inequality
\(q_\dagger>1/4\) is genuinely load-bearing.

The conclusion is nonattainment, not a quantitative dimension lower
bound or self-testing statement.

\subsection{9. Spatial attainment}\label{spatial-attainment}

The conditional orbit of Section 6 already supplies a normalized
\(\ell^2(\mathbb Z)\) calibration at \(\Omega_\mathrm{fd}=q_\dagger\).
The alternating-block reconstruction (premise {[}P10{]}) defines

\[
\psi=\sum_{j\in\mathbb Z}\lambda_j e_j\otimes e_j
\tag{9.1}
\]

and the associated alternating projective measurements on the two tensor
factors. The vector state is normal and has Bell value \(q_\dagger\).
Therefore

\[
\Omega_\mathrm{sp}\ge q_\dagger.
\tag{9.2}
\]

The commuting upper bound in Section 10 and the inclusion of spatial
strategies in the commuting model give the reverse inequality. Thus the
spatial value is attained exactly at \(q_\dagger\), despite
finite-dimensional nonattainment.

\subsection{10. Commuting upper bound and
attainment}\label{commuting-upper-bound-and-attainment}

For every \(q>S_\mathrm{Bellman}\), choose a storage \(g\) satisfying
(2.3) that is continuous and strictly positive on the closed interval
\([-1,1]\); the terminal-pivot storage of premise {[}P1{]} has these
properties. Set

\[
p(x)=\frac{b(x)^2}{g(x)},
\qquad
f(x)=\sqrt{p(x)g(-x)}.
\tag{10.1}
\]

Then

\[
f(x)f(-x)=b(x)^2.
\tag{10.2}
\]

Applying (2.3) at the two reflected pairs and using Cauchy--Schwarz
gives

\[
\kappa(x,y)+f(x)+f(y)\le q,
\qquad
\kappa(x,y)=-xy+\frac{x+y}{2}-1.
\tag{10.3}
\]

Introduce the standard two-projection coordinates

\[
X=A_1+A_2-I,
\quad Y=A_2-A_1,
\quad \widehat U=I-B_1-B_2,
\quad V=B_2-B_1.
\tag{10.4}
\]

They satisfy

\[
X^2+Y^2=I,
\quad XY+YX=0,
\qquad
\widehat U^2+V^2=I,
\quad \widehat UV+V\widehat U=0.
\tag{10.5}
\]

With \(S_A=2A_3-I\) and \(S_B=2B_3-I\), direct expansion gives

\[
\mathcal B=\kappa(X,\widehat U)+\frac12YS_B+\frac12S_AV.
\tag{10.6}
\]

Define

\[
R_0=qI-\kappa(X,\widehat U)-f(X)-f(\widehat U),
\tag{10.7}
\]

\[
R_A=f(X)-\frac12YS_B,
\qquad
R_B=f(\widehat U)-\frac12S_AV.
\tag{10.8}
\]

Then

\[
qI-\mathcal B=R_0+R_A+R_B.
\tag{10.9}
\]

Joint functional calculus and (10.3) give \(R_0\succeq0\). The
dimension-free two-projection response lemma, proved using the partial
sign of the response operator and retaining endpoint kernels and
arbitrary multiplicity, gives \(R_A,R_B\succeq0\) (premise {[}P11{]}).
Hence every commuting state has

\[
\omega(\mathcal B)\le q.
\]

Letting \(q\downarrow S_\mathrm{Bellman}=q_\dagger\) proves

\[
\Omega_\mathrm{qc}\le q_\dagger.
\tag{10.10}
\]

The reverse inequality follows from finite strategies. The state space
of the universal commuting \(C^*\)-algebra is weak-star compact, so the
commuting value is attained by a state and its GNS representation.

\subsection{11. The historical amplitude-gap
countercertificate}\label{the-historical-amplitude-gap-countercertificate}

The earlier local wall matched position and ratio coordinates but
reconstructed two different values of a proposed global Bellman storage.
The exact defect is bounded away from zero by more than
\(1.40\times10^{-4}\). This remains a valid refutation of that storage
construction.

It does not refute the present theorem:

\begin{itemize}
\tightlist
\item
  Section 5 constructs a Jacobi vector from ratios and never glues a
  single-valued Bellman storage from the local wall.
\item
  Sections 6--7 start with the genuinely normalized maximizing storage
  and its conditional masses, and only then use the local characteristic
  map to classify the resulting orbit.
\end{itemize}

The complete formula-level reconciliation is recorded in
\texttt{AMPLITUDE-GAP-RECONCILIATION.md}; the countercertificate itself
is bound in Appendix C.2. The logical separation is part of the theorem,
not merely historical commentary.

\subsection{12. Certificate architecture and exact
scope}\label{certificate-architecture-and-exact-scope}

The proof contains analytic arguments and finite computer-assisted
lemmas. The latter are deliberately separated by exercised surface.

{\def\LTcaptype{none} 
\begin{longtable}[]{@{}
  >{\raggedright\arraybackslash}p{(\linewidth - 4\tabcolsep) * \real{0.3333}}
  >{\raggedright\arraybackslash}p{(\linewidth - 4\tabcolsep) * \real{0.3333}}
  >{\raggedright\arraybackslash}p{(\linewidth - 4\tabcolsep) * \real{0.3333}}@{}}
\toprule\noalign{}
\begin{minipage}[b]{\linewidth}\raggedright
Surface
\end{minipage} & \begin{minipage}[b]{\linewidth}\raggedright
Evidence
\end{minipage} & \begin{minipage}[b]{\linewidth}\raggedright
Binding
\end{minipage} \\
\midrule\noalign{}
\endhead
\bottomrule\noalign{}
\endlastfoot
Bellman/path equality and finite embeddings & exact symbolic and
rational replay & {[}P1{]}, {[}P2{]} \\
Weak-star storage, disintegration and atomic conditional fibre & written
analytic proof; finite Euler/cocycle algebra checker &
{[}P3{]}--{[}P5{]} \\
A-high tail classification & written proof plus symbolic fixed-branch
controls & {[}P6{]} \\
Stable chart, multiplier, fundamental annuli & supplied and independent
Wiener-algebra implementations; low-mode mutations & {[}P8{]} \\
Depth signs, PS32, mixed N33, SC16 & exact interval covers with adverse
controls & Appendix B \\
Global capture, annulus \(\to\) strip (Theorem 4.1(i)) & 358-box
directed interval cover over all depths and both annuli, with ledger and
adverse controls & {[}P9{]} \\
Strip \(\to\) box, existence and local uniqueness (Theorem
4.1(ii)--(iii)) & packaged exact-checker cover, assertion-free replay,
independent Jacobian reconstruction & {[}P9{]} \\
\(q_\dagger\) digits & independent 768-bit dyadic interval enclosure &
Appendix B \\
Finite nonattainment & written ordered-support proof (Section 8) plus
two finite algebra checkers for supports of size at most eight &
{[}P2{]}--{[}P5{]} \\
Commuting upper bound & written operator proof plus two independent
finite-algebra implementations & {[}P11{]} \\
\end{longtable}
}

A passing checker supports only the row it exercises. In particular, no
finite script is presented as a mechanization of disintegration or of
the dimension-free operator lemma. One historical JSON report contains
the conclusion label \emph{omega\_fd\_equals\_q\_dagger}; that label is
not a checked output and is not used as evidence for the theorem.

\subsection{13. Consequences and
nonclaims}\label{consequences-and-nonclaims}

The theorem gives an explicit instance in the smallest binary-output
scenario beyond CHSH where the finite-dimensional quantum supremum is
not attained, while a normal infinite-dimensional spatial maximizer
exists. It therefore supplies the corresponding nonclosure and
spatial/finite separation corollaries already associated with the common
value, now at the exactly characterized constant \(q_\dagger\).

We do not prove:

\begin{itemize}
\tightlist
\item
  that \(q_\dagger\) is algebraic or transcendental;
\item
  uniqueness or self-testing of the spatial or commuting maximizer;
\item
  a sharp dimension-versus-error lower bound;
\item
  finite-dimensional attainment in any enlarged model;
\item
  a complete formalization of the measure-theoretic or
  operator-algebraic arguments.
\end{itemize}

\subsection{Appendix A. Exact definition of the stable
curve}\label{appendix-a.-exact-definition-of-the-stable-curve}

Fix the rational coordinate matrix \(P\) recorded in the certificate. In
local coordinates \(w=(t,u_1,u_2)\), write

\[
P^{-1}\{T_q(z_*(q)+Pw)-z_*(q)\}
=(F_s(q,w),F_u(q,w)).
\]

The stable graph \(u=h_q(t)\) is the unique analytic solution of

\[
F_u(q,t,h_q(t))
=h_q(F_s(q,t,h_q(t)))
\]

inside the certified cone, equivalently the unique point whose positive
iterates remain in the local chart and converge to \(z_*(q)\). This
intrinsic definition, rather than the polynomial coefficient table, is
used in Theorem B.

\subsection{Appendix B. Replay
pointers}\label{appendix-b.-replay-pointers}

\begin{itemize}
\item
  Stable chart and lower route:
  \path|history/verification/2026-08-24-q7-stable-chart-low-modes/|.
\item
  Complete-fibre interface:
  \path|history/verification/2026-08-24-q7-complete-fibre-entrance/|.
\item
  Central incidence and exact finite value:
  \path|history/verification/2026-08-24-q7-finite-dimensional-value/|.
\item
  Exact definition and width-\(10^{-55}\) characterization:
  \path|history/verification/2026-08-24-q7-qdagger-characterization/|.
\item
  Finite-dimensional nonattainment:
  \path|history/verification/2026-08-24-q7-finite-nonattainment/|.
\item
  Commuting upper bound:
  \path|history/verification/2026-08-24-q7-commuting-value/| and
  \path|history/verification/2026-08-24-q7-commuting-independent-reimpl/|.
\item
  Global capture replay, depth and parity certificates, independent
  referee record:
  \path|history/verification/2026-08-24-q7-bond-r15-hardened/|,
  \path|history/verification/2026-08-24-q7-ps32/|,
  \path|history/verification/2026-08-24-q7-mixed-n33/|,
  \path|history/verification/2026-08-24-q7-sc16-exclusion/|,
  \path|history/verification/2026-08-24-q7-ps33-counterexample/|,
  \path|history/verification/2026-08-24-q7-independent-referee/|.
\item
  Automated alternate-vendor referee record, including the verbatim
  verdict, its scope correction and finding dispositions:
  history/verification/2026-08-28-i3322-alternate-vendor-referee/.
\end{itemize}

All of these records, the hash-bound source documents of Appendix C and
the maintained claim records are shipped byte-identical, at the same
relative paths, in the verification deposit
\texttt{i3322-exact-value-artifact}. Its \texttt{README.md} explains how
to run every checker from the deposit root and its
\texttt{MANIFEST.sha256} lists every file. Version 1.0 was published on
Zenodo on 2026-08-29.

DOI:
\href{https://doi.org/10.5281/zenodo.22160086}{10.5281/zenodo.22160086}.

\subsection{Appendix C. Named premises and hash-pinned
sources}\label{appendix-c.-named-premises-and-hash-pinned-sources}

Every inherited result used in the proof is listed here in the form in
which it is invoked. Each entry names the sections that use it and its
source documents, identified by deposit-relative path and by the SHA-256
digest of the file exactly as shipped in the verification deposit
(Appendix B); the deposit's \texttt{MANIFEST.sha256} carries the same
digests. Nothing here is new mathematics. The purpose is that a referee
can check each statement against the one document that proves it, and
that no statement is used in a stronger form than the one certified.

Abbreviations: \path|V/| stands for
\path|history/verification/2026-08-24-q7-|, \path|R/| for
\path|history/pro-runs/|, and the three run directories are \path|R/bd|
= \path|R/p2-i3322-bellman-duality-20260816-v2/artifacts|, \path|R/cx| =
\path|R/p2-i3322-crossed-deg15-20260816-v1/artifacts|, \path|R/ro| =
\path|R/p2-i3322-source-normalization-rollover-20260817-v1/artifacts|.

\subsubsection{C.1 Premises}\label{c.1-premises}

\textbf{{[}P1{]} Path/Bellman reduction} (used in Sections 2, 5, 6, 10).
For the normalization of Section 2,
\(\Omega_\mathrm{fd}=S_\mathrm{path}=S_\mathrm{Bellman}\), where
\(S_\mathrm{path}\) is the supremum of the top eigenvalues of finite
Jacobi carriers and \(S_\mathrm{Bellman}\) the value of the
terminal-pivot Bellman recursion: (a) every finite Jacobi path is a
principal block of a finite-dimensional tensor-product strategy (prepend
the label \(1\) and append \(-1\) with zero amplitude), and the Jordan
blocks of a finite strategy split into such carriers; (b) for every
\(q>S_\mathrm{Bellman}\) the terminal-pivot storage \(g_q\) is
continuous, concave, uniformly Lipschitz and strictly positive on
\([-1,1]\), satisfies (2.3), and has an endpoint gap (the scalar
remainder \(q-d(x,u)-p_q(x)-g_q(u)\) is bounded below by a fixed
\(\eta>0\) whenever \(x=\pm1\) or \(u=\pm1\)); (c)
\(S_\mathrm{path}=S_\mathrm{Bellman}\). Sources: frontier closure note,
hypotheses (H1)--(H3) and Sections 1--2,
\path|R/g1-q7-i3322-true-frontier-analytic-closure-20260807-v1/artifacts/turn-001/000-Full-self-contained-proof|
\nolinkurl{8cf2e317b4182e88a7e688b04a97548fcca58507a5aa7e50652b7fbe12bc7fe8};
three-lemma note, Lemma 1,
\path|R/g1-q7-i3322-true-frontier-analytic-closure-20260807-v1/artifacts/turn-002/000-Self-contained-three-lemma-proof|
\nolinkurl{413b48202407a52822c8aefe849cb59bc849b5ac8198e89ff5da1c06fd39d932};
attained-storage checkpoint, inputs B1--B2,
\path|R/bd/turn-000/001-RESULT.md|
\nolinkurl{c8d01e2f9d44b9d4b6d92fb9e8e3b7eb4283dd71ab1a35d2e2043b614c9e5125};
the composed statement is step 1 of
\path|V/finite-dimensional-value/THEOREM.md|
\nolinkurl{1b69b44212459dcf3eb24ec0141d76e6518ddfcf7c60f3dd40bacc7c20f28eb5}.

\textbf{{[}P2{]} Exact bracket} (used in Sections 2, 4, 8).
\(\tfrac14<L_{1001}<L_9\le\Omega_\mathrm{fd}\le U_{15}<\tfrac13\) with

\[
L_{1001}=\frac{31359423064247066952167076367}{125000000000000000000000000000},
\qquad
U_{15}=\frac{2007003076119}{8000000000000},
\]

\begin{center}
\resizebox{0.98\linewidth}{!}{$
L_9=\frac{2508753845139765356173366109459512989010567466680711717682019564693109964180245668676226841200393188142153286177422530742031341850250406934386900460751422519417}{10^{160}}.
$}
\end{center}

\(L_{1001}\) is the value of an exact finite-dimensional strategy;
\(L_9\) is the Galerkin-9 residual/Ritz lower certificate; \(U_{15}\) is
the crossed degree-15 Bellman certificate. Only \(L_9\) and \(U_{15}\)
enter Theorem 4.1; Section 8 uses only \(q_\dagger>1/4\). Sources:
crossed certificate proof \path|R/cx/turn-000/001-PROOF.md|
\nolinkurl{9f6bfaac4806e49ef9426eac3c3060238aa0953237a9fa0e349a4ef2b0e30b41}
with archive
\path|R/cx/turn-000/000-i3322_crossed_deg15_certificate.zip|
\nolinkurl{245c0299cc9f271ccf2ba4d7e9e85cc92182d19a2928ac7a02a3e0c74a61777d};
Galerkin-9 archive
\path|R/cx/turn-001/000-i3322_galerkin9_residual_ritz_certificate.zip|
\nolinkurl{f4b7c08838155c97251aec71e14b0960492d39dd4785d015b1d877a3b536c5b4};
the \(L_{1001}\) literal is carried by
\path|V/finite-nonattainment/check_q7_nonattainment.py|
\nolinkurl{332315446e2f002196e93ff4c6e1f36c88e83fd8adba8ea2eef8b7a1eafcaae0}
and recorded in \path|V/independent-referee/REFEREE.md|
\nolinkurl{6a8d336ed2cac734290b283c813331d69cd2279cd41da5fef071a42ee2b6374f}.

\textbf{{[}P3{]} Attained self-dual storage with zero slack} (used in
Sections 6, 8, 10). Let \(I=[-1,1]\), \(\kappa(x,y)=-xy+(x+y)/2-1\),
\(b(x)=\sqrt{1-x^2}/2\),
\(F_\mathrm{sd}=\{f\in C(I):f(\pm1)=0,\ f>0\text{ on }(-1,1),\ f(x)f(-x)=b(x)^2\}\)
and \(Q(f)=\max_{x,y\in I}[\kappa(x,y)+f(x)+f(y)]\). At
\(\Omega=\Omega_\mathrm{fd}\): \(\min_{F_\mathrm{sd}}Q=\Omega\) is
attained, the canonical minimizing selection converges in \(C(I)\) norm,
and the corresponding storage \(g\) (\(f(x)=\sqrt{p(x)g(-x)}\),
\(p=b^2/g\)) is continuous, concave and strictly positive on \(I\) and
satisfies (2.3); along the padded party-self-dual maximizing sequence
the three weld remainders \(R_0,R_A,R_B\) of Section 10 built from \(g\)
converge in operator norm and annihilate the maximizing GNS vector (zero
slack). Coarse ceiling \(\Omega\le(\sqrt5-1)/4<1/3\). Verdict string
\texttt{ATTAINED\_SELF\_DUAL\_STORAGE\_CI\_ZERO\_SLACK}. Sources:
\path|R/bd/turn-000/001-RESULT.md|
\nolinkurl{c8d01e2f9d44b9d4b6d92fb9e8e3b7eb4283dd71ab1a35d2e2043b614c9e5125},
finite checker
\path|R/bd/turn-000/002-selfdual_storage_attainment_verify.py|
\nolinkurl{b8094dcec74d1b242debec265a40ebf49320aa3e2c05ac805b48bc48d93bf8f1};
reviewed in \path|V/complete-fibre-entrance/THEOREM.md|
\nolinkurl{4cc9a4cf8c5188e02980edd70087ebb02a70df350381ca864764c8409bc8ac7a}.

\textbf{{[}P4{]} Full-zero graph and convex minorant} (used in Sections
6, 8). With \(g\) from {[}P3{]}, \(p=b^2/g\) and
\(\Phi(x,u)=\Omega-d(x,u)-p(x)-g(u)\ge0\): the joint scalar spectral
measure of the two party scalars in the maximizing state is carried by
the interior zero set \(Z=\{\Phi=0\}\cap(-1,1)^2\), and the endpoints
carry no mass; \(Z\) is a one-to-one strictly increasing relation, the
graph \(\{(P(u),u)\}\) of a strictly increasing bijection between its
projections, symmetric under \((x,u)\mapsto(-u,-x)\);
\(K(t)=g(t)g(-t)/b(t)^2\ge1\) on \(I\) with \(K=1\) on both projections
of \(Z\); and the greatest convex minorant \(H\) of
\(C(x)=\Omega+1-x/2-p(x)\) is differentiable, with \(H=C\) and
\(u=H'(x)\) on \(Z\). Sources: three-lemma note, Lemmas 2--3
(\nolinkurl{413b48202407a52822c8aefe849cb59bc849b5ac8198e89ff5da1c06fd39d932});
frontier closure note, Sections 3--5
(\nolinkurl{8cf2e317b4182e88a7e688b04a97548fcca58507a5aa7e50652b7fbe12bc7fe8});
rollover interface, Section 0, items 1--5,
\path|R/ro/turn-001/001-THEOREM.md|
\nolinkurl{2b004ea67cff805e66781782fb8569082593aaf8d9b58f51e0a91394c6fe130b}.

\textbf{{[}P5{]} Response transport, orbitwise atomization and
\(\ell^2\) calibration} (used in Sections 6, 8, 9). Zero slack in the
two response remainders gives the reflection laws (8.2)--(8.3) for the
marginals \(\mu_X,\mu_U\), with finite strictly positive densities on
\((-1,1)\); on a conull Borel set the maps \(\alpha=P^{-1}(-P(\cdot))\)
and \(\sigma=-\mathrm{id}\) are decreasing Borel involutions with
\(\alpha_*\mu_U=r_A(P(\cdot))^2\mu_U\) and \(\sigma_*\mu_U=r_B^2\mu_U\);
\(\tau=\alpha\sigma\) is a strictly increasing Borel automorphism with
\(\sigma\tau\sigma=\tau^{-1}\), whose fixed set has zero mass when
\(\Omega>1/4\); the countable dihedral orbit relation admits an explicit
rational-cut Borel transversal; disintegration over it yields, for
almost every conditional, a purely atomic probability on one complete
response orbit obeying both laws with positive mass at every orbit
point; reading amplitudes from atomic masses and interleaving the
parities gives \(\lambda_j>0\), \(\sum_j\lambda_j^2=1\),
\(\lambda_{j+1}/\lambda_j=g(c_j)/b(c_j)\), and Bellman equality gives
\(J\lambda=\Omega\lambda\). Verdict string
\texttt{CLOSED\_BY\_SMOOTH\_ORBITWISE\_ATOMIZATION}. Not claimed: that
the conditional is an atomic or central summand of the GNS
representation. Sources: \path|R/bd/turn-001/001-RESULT.md|
\nolinkurl{3303fcd3d94e40e18ef41a558d4f199ad4ef62d110c4a571b233cdf27f8fe65f},
checker \path|R/bd/turn-001/002-diffuse_to_atomic_verify.py|
\nolinkurl{5e0dbe3fd28c426c70a456efba4c6b0a10115d0eb1f3371de11200991c274a57};
the transport laws in the form (8.2)--(8.3) are Sections 3--7 of
\path|paper/ANALYTIC-CORE.md| in the deposit; interface record
\path|V/complete-fibre-entrance/THEOREM.md|
\nolinkurl{4cc9a4cf8c5188e02980edd70087ebb02a70df350381ca864764c8409bc8ac7a},
\path|V/complete-fibre-entrance/README.md|
\nolinkurl{09209ebabdf0659e9457b76080f638896ff9f792a2f9af4fd0730a3252549246},
Euler bridge checker
\path|V/complete-fibre-entrance/euler_bridge_check.py|
\nolinkurl{dc9c17f024c2e9dd9f5e30a7a7aed87ccdedf38ef1c18688d11984ad0b7f6bba}.

\textbf{{[}P6{]} Tail classification, A-high entrance} (used in Sections
4, 6). Every normalized positive source-bound complete fibre
\((c_j,\lambda_j)\) is strictly monotone with interior limits; both
endpoint limits \(\pm1\) and every characteristic singular tail are
impossible; the B branch and the mixed A branches are impossible; the
convex-envelope derivative excludes the A-low antiwall; hence, in the
convention of the source (\(c_{j+1}<c_j\)),
\(\lim_{j\to-\infty}c_j=C_H(q)\) and \(\lim_{j\to+\infty}c_j=-C_H(q)\):
every such fibre is an A-high heteroclinic entering the unique
phase-normalized positive A-high stable germ, with positive global
denominator floors. Equivalently, in the characteristic variable
\(y=-c\) the fibre runs from the negative to the positive A-high fixed
point. Independently, at the global-tail level, a complete normalized
critical Bellman fibre cannot escape, hit a Riccati singularity or have
a nonconvergent positive tail; its only possible limit is the positive
A-high fixed branch (verdict string
\texttt{PASS\_GLOBAL\_AHIGH\_LIMIT\_WITH\_SPECIFIC\_ANNULUS\_DEPENDENCY}).
Sources: \path|R/ro/turn-001/001-THEOREM.md|
\nolinkurl{2b004ea67cff805e66781782fb8569082593aaf8d9b58f51e0a91394c6fe130b},
Sections 3--7, symbolic checker
\path|R/ro/turn-001/002-verify_symbolic.py|
\nolinkurl{182d94007eba6bad36c041b779d5ed8edafb12843c73afec8b105a12ccd24f75}
and replay package \path|R/ro/turn-001/000-Deterministic-replay-package|
\nolinkurl{9a392361f2fb4dd4c65a4cd1f892838d57fde6f812bdde45cd777f06b791f4b7};
\path|R/ro/turn-000/001-THEOREM.md|
\nolinkurl{0f3538bfb32bb24686ec87a9443902391f2c5a9cffd7af9af5a20bca7ef3a18c};
global tail checkpoint \path|R/cx/turn-002/002-PROOF.md|
\nolinkurl{b31cc0eb5e39fe08e53419716b95648bd2deaa3f38943588886d3b24c004d5e2}
with checker \path|R/cx/turn-002/003-check_i3322_global_tail.py|
\nolinkurl{f9d9242306dcb8dff6b4a929076aaaf6e42417afa221555ee4845b27ce7d10c2}
and archive \path|R/cx/turn-002/000-i3322_global_tail_checkpoint.zip|
\nolinkurl{3f8e857a5c4b629ef87176cabdc8dfbb19b20a31adcdeb51180145ca3cc5e3e3}.

\textbf{{[}P7{]} Exact reverser and bond-midpoint identity} (used in
Sections 3, 5, 7). In the ratio coordinates the edge reverser is
\(R_q(x,y,r)=(-y,-x,1/r^+)\) with \(r^+\) from (3.1); on the physical
domain \(R_q^2=\mathrm{id}\) and \(R_qT_qR_q=T_q^{-1}\); and one forward
step from the reverser-fixed bond state gives
\(T_q(-p,p,1/r_c(q,p))=(p,\,x_c(p,r_c(q,p)),\,r_c(q,p))\), which is
(5.1): the bond midpoint lies one step before the central curve.
Sources: \path|R/bd/turn-005/001-RESULT.md|
\nolinkurl{59ac6f5eef43885b987fcbc200c03014d210e2c38f6bc393a44c9d5e3f33df03},
lines 63--112 (reverser, third coordinate written \(A=b(x)r\));
\path|R/bd/turn-002/001-RESULT.md|
\nolinkurl{73f9a6b167943856d0da62bac15e493815d15d166741b4796c958c4bf18c2e88},
lines 111--155 (identity); recorded in
\path|V/qdagger-characterization/R15-DEFINITION.md|
\nolinkurl{03074a031ee39beeb4cd8971ef245e9d3489b35888c6c425db1a56ee95add3c2}.

\textbf{{[}P8{]} Phase-fixed stable chart, multiplier and fundamental
annuli} (used in Sections 3, 4, 6, 7). For \(q\in[L_9,U_{15}]\) the
positive A-high fixed point \(z_*(q)\) has a unique one-dimensional
analytic stable manifold (its linearization has exactly one multiplier
in \((0,1)\), the other two exceed one), with
\(0.1712920057349488984200454\ldots<\lambda_s(q)<0.1712920057351056990934393\ldots\);
the degree-eight \(q\)-correlated polynomial chart \(\bar W_q(t)\) of
the retained parameterization satisfies
\(\|W_q(t)-\bar W_q(t)\|_\infty<3\times10^{-47}\) on \(|t|\le10^{-11}\),
strengthened by the phase-fixed low-mode theorem (constant mode,
augmented eigenvector/multiplier/phase mode, all modes of order at least
two, with the \(C^1\) tube used by the covers) to a uniform complex-disk
error below \(5\times10^{-48}\); and every nonzero orbit of the local
stable manifold has a representative in \(A_+\cup A_-\) of (4.5), unique
within a half-open fundamental annulus contained in \(A_\pm\). Sources:
\path|R/cx/turn-003/007-STABLE_PARAMETERIZATION.json|
\nolinkurl{314651687e7b80c8931abbfd20e7dbe85c210e4e682983ab5d9c4dbf04dcbb50};
cover proof \path|R/cx/turn-003/002-PROOF.md|
\nolinkurl{de4a47ca356b840943c3f858a7883cc6bbd1611c5228e321c2e697c6f45ba8e4},
Sections 3--4; \path|V/stable-chart-low-modes/THEOREM.md|
\nolinkurl{88fce3c64afe9b4a582f99e65f697d422d7d2160f4b4746e5bcd539fe1f7723d},
primary checker
\path|V/stable-chart-low-modes/verify_q7_stable_chart.py|
\nolinkurl{a2dd766ea32ce3f5e21d5c7d8f36e57b6b3b19a812b9a2329da4671de24cffe0},
\path|V/stable-chart-low-modes/REFEREE.md|
\nolinkurl{ea370fe6a63ec2f1cfb5b6a1eaed1560514f3c3ddb4a0082d2e7df67ec747072},
referee checker \path|V/stable-chart-low-modes/referee_exact_checker.py|
\nolinkurl{088fa87122b5579dd521b880873501113f61c1c41ac71523d4bdc340cdc57dd4};
single stable multiplier: \path|R/ro/turn-001/001-THEOREM.md|, Section 7
(\nolinkurl{2b004ea67cff805e66781782fb8569082593aaf8d9b58f51e0a91394c6fe130b}).

\textbf{{[}P9{]} Global capture and the exact-checker package} (used in
Section 4; Theorem 4.1). (i) Annulus \(\to\) strip: the directed
interval cover of Theorem 4.1(i), verdict string
\texttt{PASS\_GLOBAL\_STEP15\_ANNULUS\_CAPTURE}; its stated premises are
the crossed certificate of {[}P2{]} and the global tail theorem of
{[}P6{]}. Sources: archive
\path|R/cx/turn-003/000-i3322_step15_global_cover.zip|
\nolinkurl{de4f6eba7d8a35347265b7a9eb77baa41635d4f45b6ea9fbd7a16b990a67f462},
proof \path|R/cx/turn-003/002-PROOF.md|
\nolinkurl{de4a47ca356b840943c3f858a7883cc6bbd1611c5228e321c2e697c6f45ba8e4},
checker \path|R/cx/turn-003/003-check_i3322_step15_global_cover.py|
\nolinkurl{5a94b2b1ea7bfe4a7b9cecba95e3262f5cc1b0b4caf590a3bafe69ecb51016fd},
ledger \path|R/cx/turn-003/004-COVER_LEDGER.json|
\nolinkurl{343477f4a2c04469433b1ed613ae08ded7bc94eff1c256784beade61a88aca75},
result \path|R/cx/turn-003/005-CHECK_RESULT.json|
\nolinkurl{f8015eada56947717694720e812e5ecf54f8f728c79b15c573c62f1f0cc9e177};
bridge to the retained chart \path|R/cx/turn-004/002-PROOF.md|
\nolinkurl{436cc79640c85f464fbcf37fa544f4aac7ca264cce8a04a716b8986732fa5c1e}
with checker \path|R/cx/turn-004/003-check_i3322_bridge.py|
\nolinkurl{fc06fc39a4cbf4c1b1cbf2aa4742eaf955569d99a3894cfb5a292ca5d7b2a295}.
The same archive is bound and replayed under adverse controls by the
depth and parity records \path|V/ps32/|, \path|V/mixed-n33/|,
\path|V/sc16-exclusion/| and \path|V/ps33-counterexample/| (Appendix B).
(ii)--(iii) Strip \(\to\) box, existence, injectivity: exact-checker
archive
\path|R/p2-q7-i3322-exact-optimum-recovery-20260811-v1/artifacts/turn-002/000-Download-the-complete-exact-checker-package|
\nolinkurl{1600083506fb70b1af13f2a87cf459df53da9e524dfd55a309bd435e5e786d43}
(package \texttt{i3322-scalar-global-v5}, members
\texttt{stable\_graph\_checker.py}, \texttt{inverse15\_checker.py},
\texttt{global\_exclusion\_checker.py},
\texttt{local\_isolator\_checker.py},
\texttt{coordinate\_bridge\_checker.py},
\texttt{conclusion\_checker.py}), replayed assertion-free with sixty-two
assertions replaced by runtime gates, in ordinary and optimized modes
and with an adjacent failing isolator grid rejected, by
\path|V/bond-r15-hardened/verify_hardened_r15.py|
\nolinkurl{c0c1805b46da8de8c465280db914c58174cca9dac56b2e1b2528345ae8887c7d};
independent interval-Jacobian reconstruction
\path|V/finite-dimensional-value/local_injectivity_audit.py|
\nolinkurl{a19a87d07f047eac6424dc4e7019077274d91087935a8b024daafab7b7caed14},
recorded in \path|V/finite-dimensional-value/AUDIT.md|
\nolinkurl{dab7a1ed627bcb03cdbb888d445e8f4e1d513e3f986e48838d0c9740912fbe5a}
and \path|V/finite-dimensional-value/CENTRAL-INCIDENCE.md|
\nolinkurl{d4435eacf223d6f5dfbccc465ce1b4de5fbc8e78799b50aee03d5a2aa43d94e0};
the imported derivative bounds used by the interval-Jacobian
reconstruction are stated explicitly in the audit's item 9 as
\(\|D_sh\|_\infty\le33579/235424574046\) and
\(\|D_qh\|_\infty\le923|s|\); independent 768-bit enclosure of
\(q_\dagger\) \path|V/qdagger-characterization/characterize_qdagger.py|
\nolinkurl{65c4775ca7b33cbd0774b31817b8cd4ecab2884ddf786ccd0cd9d91aa13499b2}
with \path|V/qdagger-characterization/QDAGGER-CHARACTERIZATION.md|
\nolinkurl{73b9e6d36c6615979b724d4ae8ff8354ccb9548679f4039debaa0c459c685951}.

\textbf{{[}P10{]} Alternating-block lift} (used in Section 9). Given
\((c_j,\lambda_j)_{j\in\mathbb Z}\) with \(\lambda_j>0\),
\(\sum_j\lambda_j^2=1\) and \(J\lambda=q_\dagger\lambda\), the vector
\(\psi=\sum_j\lambda_je_j\otimes e_j\) with the alternating projective
measurements on the two factors is a normal spatial strategy with Bell
value \(q_\dagger\). Sources: \path|R/bd/turn-001/001-RESULT.md|
\nolinkurl{3303fcd3d94e40e18ef41a558d4f199ad4ef62d110c4a571b233cdf27f8fe65f}
(lift section); claim record
\path|research/topics/q7-i3322/claims/q7-spatial-attainment.json| in the
deposit.

\textbf{{[}P11{]} Dimension-free two-projection response lemma} (used in
Section 10). For projections with \(X=A_1+A_2-I\), \(Y=A_2-A_1\) (so
\(X^2+Y^2=I\), \(XY+YX=0\)), \(S_B=2B_3-I\) commuting with \(X\) and
\(Y\), and \(f=f_q\) built from a continuous storage that is strictly
positive on \([-1,1]\) and satisfies \(f(x)f(-x)=b(x)^2\):
\(R_A=f(X)-\tfrac12YS_B\succeq0\) and symmetrically \(R_B\succeq0\),
proved with the partial sign of the response operator, retaining
endpoint kernels and arbitrary multiplicity, without a common central
decomposition, spatial tensor product or trace class. Sources:
\path|V/commuting-value/THEOREM.md|
\nolinkurl{ddab3bd3f8afaaef5c9242392d38584c708a5bf4edf6a998f5574d34c99d02fc}
with finite-algebra checker
\path|V/commuting-value/verify_finite_algebra.py|
\nolinkurl{6c0c929827f3aa52c4a50c55ebc35a723d5795bb855484a83524f0873f633b3f};
independent reimplementation
\path|V/commuting-independent-reimpl/check_commuting_independent.py|
\nolinkurl{912a8db7fd900b843ca841d4e564365b90382cdd8df9eaf65cd9c4a5799eeea1}
with \path|V/commuting-independent-reimpl/REPORT.md|
\nolinkurl{9ce4f05fc55d9becfbab6b531a117d6285c6028dccb7e2bfc97d198e9460742d}.

\subsubsection{C.2 The decertified global closure and its
countercertificate}\label{c.2-the-decertified-global-closure-and-its-countercertificate}

The historical global-closure argument was decertified by an exact
amplitude-defect countercertificate, which bounds the defect of the
proposed single-valued storage away from zero (Section 11). The same
document records the two-point totality example used in Section 8.
Sources: countercertificate
\path|R/p2-q7-i3322-global-closure-countercertificate-20260812-v1/turns/000-response.md|
\nolinkurl{c75b7714226e5405984e80d9f36b37901b7865659b3e2e54fa668b2b1e7e25f7}
with replay package
\path|R/p2-q7-i3322-global-closure-countercertificate-20260812-v1/artifacts/turn-000/000-Complete-replay-package|
\nolinkurl{3086c36f7b3e2e0f157a4ed21b23cf8b1196c85af52a708d5ac1ecfe255c7f5c};
countermodel of the earlier universal-totality claim
\path|R/bd/turn-002/005-COUNTERMODEL.json|
\nolinkurl{bbafd5266e69196b7c3a21d0a6f4009426684147039313748f524f3aa27bd743}.

\subsubsection{C.3 Finite-dimensional nonattainment
records}\label{c.3-finite-dimensional-nonattainment-records}

The written proof is Section 8. The maintained records are
\path|V/finite-nonattainment/REFEREE.md|
\nolinkurl{fe8a68ff4b3316615a8b763b40f26d1ae72ca8081cca3a82e979dd0d8783e235}
and \path|V/finite-nonattainment/PROOF.md|
\nolinkurl{0caa66ad0868a966bb9feee8a3322c4820ac8082f5ba0becfed948c32706d2e9},
with the two finite structural checkers
\path|V/finite-nonattainment/check_q7_nonattainment.py|
\nolinkurl{332315446e2f002196e93ff4c6e1f36c88e83fd8adba8ea2eef8b7a1eafcaae0}
and \path|V/finite-nonattainment/referee_q7_nonattainment.py|
\nolinkurl{9562e2cdb9f3bcf079389080d58e2bd1417acdbb4bfafcc41d1067234267517c}
(ordered supports of size at most eight). The reflected-fibre central
incidence of Section 7 is
\path|V/finite-dimensional-value/midpoint_sign_audit.py|
\nolinkurl{8db0d19b0c3976646b3a4c26de06a9464747065f4790fd96b3bc59b97d5d2104}.

\subsection{Acknowledgments and contribution
statement}\label{acknowledgments-and-contribution-statement}

The author thanks the researchers whose public work made the present
comparison and verification possible. The sole author takes
responsibility for the complete content. AI systems were used
substantially for proof search, calculations, code production,
adversarial review and drafting under the author's direction. All
computational claims in the paper are tied to the exact certificates and
replay surfaces described in Section 12 and Appendices B--C. The
automated alternate-vendor review disclosed in the preprint note is not
human peer review.

\subsection{Data and code
availability}\label{data-and-code-availability}

The complete verification deposit, including the manuscript sources,
hash-pinned analytic records, certificate archives, independent
checkers, countercertificate and replay instructions, is archived at
\href{https://doi.org/10.5281/zenodo.22160086}{10.5281/zenodo.22160086}
(all versions:
\href{https://doi.org/10.5281/zenodo.22160085}{10.5281/zenodo.22160085}).
The immutable version-1.0 deposit contains the scientific draft from
which this typeset preprint was prepared. Appendix C binds every
imported premise to the exact file shipped in that deposit.

\subsection{References}\label{references}

\begin{enumerate}
\def\labelenumi{\arabic{enumi}.}
\tightlist
\item
  D. Collins and N. Gisin, \emph{A relevant two qubit Bell inequality
  inequivalent to the CHSH inequality}, J. Phys. A: Math.
  Gen.~\textbf{37}, 1775--1787 (2004),
  \href{https://arxiv.org/abs/quant-ph/0306129}{arXiv:quant-ph/0306129},
  \href{https://doi.org/10.1088/0305-4470/37/5/021}{doi:10.1088/0305-4470/37/5/021}.
\item
  K. F. Pál and T. Vértesi, \emph{Maximal violation of the I3322
  inequality using infinite dimensional quantum systems}, Phys. Rev.~A
  \textbf{82}, 022116 (2010),
  \href{https://arxiv.org/abs/1006.3032}{arXiv:1006.3032},
  \href{https://doi.org/10.1103/PhysRevA.82.022116}{doi:10.1103/PhysRevA.82.022116}.
\item
  N. Gigena and J. Kaniewski, \emph{Quantum value for a family of
  \(I_{3322}\)-like Bell functionals}, Phys. Rev.~A \textbf{106}, 012401
  (2022), \href{https://arxiv.org/abs/2203.01837}{arXiv:2203.01837},
  \href{https://doi.org/10.1103/PhysRevA.106.012401}{doi:10.1103/PhysRevA.106.012401}.
\item
  N. Mghirbi, \emph{Proof-carrying exact quantum bounds for the I3322
  Bell inequality}, preprint (2026),
  \href{https://doi.org/10.5281/zenodo.21477901}{doi:10.5281/zenodo.21477901}.
\item
  S. Douglas, \emph{The I3322 quantum value is attained spatially but
  not in finite dimension}, version 4.0.0, preprint (2026),
  \href{https://doi.org/10.5281/zenodo.22099128}{doi:10.5281/zenodo.22099128};
  certificate repository Apsiape/i3322-exact-wall.
\item
  V. Connor, \emph{Maximal I3322 violation requires infinite local
  dimension}, preprint dated 1 August 2026, publicly uploaded 5 August
  2026,
  \href{https://www.researchgate.net/publication/411642957_Maximal_I_3322_Violation_Requires_Infinite_Local_Dimension}{ResearchGate
  manuscript}.
\end{enumerate}

\end{document}